# Transitioning ECP Software Technology into a Foundation for Sustainable Research Software

**Gregory R. Watson**[a], Addi Malviya-Thakur[a], Daniel S. Katz[b], Elaine M. Raybourn[c], Bill Hoffman[d], Dana Robinson[e], John Kellerman[f], Clark Roundy[f]

## Abstract

Research software plays a crucial role in advancing scientific knowledge, but ensuring its sustainability, maintainability, and long-term viability is an ongoing challenge. The Sustainable Research Software Institute (SRSI) Model has been designed to address the concerns, and presents a comprehensive framework designed to promote sustainable practices in the research software community. However the SRSI Model does not address the transitional requirements for the Exascale Computing Project (ECP) Software Technology (ECP-ST) focus area specifically. This white paper provides an overview and detailed description of how ECP-ST will transition into the SRSI in a compressed time frame that a) meets the needs of the ECP end-of-technical-activities deadline; and b) ensures the continuity of the sustainability efforts that are already underway.

[a] Oak Ridge National Laboratory, Oak Ridge, TN 37380
[b] University of Illinois, Urbana, IL 61801
[c] Sandia National Laboratories, Albuquerque, NM 87185
[d] Kitware Inc., Clifton Park, NY 12065
[e] The HDF Group, Champaign, IL 61820
[f] Eclipse.org Foundation Inc., Portland OR 97206

# 1. Background

The Exascale Computing Project (ECP) has been a great success, and has resulted in a software ecosystem for exascale computing that has not only enabled significant science to be undertaken, but that will be essential for realizing the future of scientific endeavor in the post-exascale era. However, with the announced completion of technical work on ECP set for December 31, 2023[1] and funding for around 1000 people set to end, a significant inflection point is about to be reached. As one of six seedling projects funded by the Department of Energy (DOE) tasked with examining how to preserve and enhance this sociotechnical ecosystem of software and people, our open, neutral, community-driven approach offers a fresh alternative to traditional funding models that have been employed in the past. In this paper, we describe an interim model for transitioning ECP Software Technology projects and teams to our proposed Sustainable Research Software Institute (SRSI), while at the same time beginning to create a post-exascale software sustainability foundation organization. A complete description of our SRSI model available in the white paper "[An Open Community-Driven Model For Sustainable Research Software: Sustainable Research Software Institute](#)".

The ECP Software Technology (ECP-ST) focus area[2] is one of the three main focus areas of the ECP. It is responsible for developing the critical underlying technologies that enable applications to be developed for exascale systems and is responsible for the Extreme-scale Scientific Software Stack (E4S) effort. It comprises approximately 70 products from 35 subprojects that are developed collaboratively by 9 DOE labs and 22 universities, as well as a number of vendors, and is supported by numerous open source contributors. ECP-ST staff also participate in a variety of standards efforts, and have close collaborative arrangements with vendors and third-party software companies. ECP-ST is composed of six communities (known as L3 technical areas): programming models & runtimes; development tools; math libraries; data and visualization; software ecosystem; and NNSA. The number of subprojects within these communities ranges from three (ecosystem) to nine (programming models & runtimes). ECP-ST products are delivered primarily as source code in open source version control platforms such as GitHub and GitLab. E4S is increasingly becoming the primary conduit for access to ECP-ST products, and has the stated policy of accepting contributions from non-DOE teams. It provides a set of community policies that serve as membership criteria for projects to become part of the E4S ecosystem.

---

[1] https://www.hpcwire.com/2023/05/31/lori-diachin-to-lead-the-exascale-computing-project-as-it-nears-final-milestones/#
[2] https://info.ornl.gov/sites/publications/Files/Pub184790.pdf

In anticipation of the impending end of ECP, this transition plan has been developed specifically for ECP-ST to ensure the continuity of funding for ECP-ST products for sustainability related efforts until it can be managed by the future software sustainability foundation organization, SRSI. A successful transition plan will leverage the buy-in of the current ST leadership, facilities, national laboratories, academic institutions, and others involved in the ECP. This plan ensures that the current infrastructure support that the ECP organization currently provides that is critical to ST teams is carefully considered during a transition. The transition will be staged so as to minimally disrupt the existing ECP-ST structure, while ensuring the ongoing operation of necessary infrastructure, and that vendor and other important interactions are able to continue uninterrupted.

## 2. Organization Establishment

The first action undertaken will be to establish an organizational structure for SRSI so that funding arrangements with participating organizations can be instituted, contracts to preserve existing infrastructure put in place, and that disruption to project activities can be kept to an absolute minimum.Discussions are currently underway with both the Linux Foundation and NumFOCUS to provide a simplified path for realizing the SRSI organization. The primary determinant will be the efficiency in establishing a legal entity and the flexibility of the arrangement in preserving the principles embodied in the SRSI Model. In either case, SRSI will require a corporate legal structure, bylaws, initial staff, and an initial board of directors or the equivalent. SRSI will also be provided with a draft policy describing the participatory community program structure for the organization. In parallel with this, a set of bylaws for the SRSI Foundation will be provided so that funding arrangements with participating organizations can be established and funds can accepted and distributed on behalf of granting bodies.

Once the organizational structures and staffing are in place, SRSI will transition existing testing and CI/CD infrastructure by establishing a 12-month contract with Paratools to preserve the current arrangement being employed by ECP-ST. Additionally, and if needed, a 12-month contract will be established with a vendor (such as Kitware) to provide contract management services and sustainability support services for SRSI. At the conclusion of the 12-month period, these contracts will be put out for a competitive bid.

## 3. Community Program Transition

In SRSI, community programs are used to enable a number of projects to work together to build a collective ecosystem. Each community program needs to provide a governance structure and meritocracy for sub-communities and projects. During the transition phase, SRSI will work with the ECP-ST leadership to transition to one or more community programs[3]. If an ECP organizational unit transitioning to a community program wishes to keep the existing organizational structure intact, each L3 technical area could continue to operate as sub-communities of the community program. SRSI will work with each of the L3 technical areas to determine which projects and supporting organizations providing technical resources wish to transition to SRSI. In either case, the current L3 technical area leads can continue in their current roles, or these roles can be transferred to different personnel. The community program lead would be determined by the overall community via a community election process. As part of the transition to SRSI, each participating project will also need to undertake a sustainability assessment in the form of a sustainability scorecard. Existing assessments of a project's sustainability needs, would be leveraged if available rather than attempt to recreate it. The SRSI Foundation will then establish CRADAs and other funding arrangements with the identified participating organizations.

It may also be attractive to DOE if select "nuggets" from the other seedling projects are used to pull together the best organizational fit for sustaining ECP-ST projects. SRSI would welcome the opportunity to collaborate with other seedling proposers to apply the best approaches for DOE SC and the ECP-ST to the planning and execution of SRSI and the SRSI Foundation.

# 4. Transitional Funding Arrangements

Funding required for the initial ramp-up and operation of SRSI, as well as funding for sustainability activities on the ECP-ST community program(s) will be provided by DOE. This funding will be augmented as the SRSI expands its scope to other agencies, organizations, and communities. It will not be relying exclusively on DOE funding to operate, other than during the transitional period, which is currently planned to be twelve months.

## 4.1 Operational Funding: SRSI

Initially, DOE will provide SRSI membership fees to each of the supporting organizations (e.g., national laboratories) identified during the transition of L3 technical areas to community programs. These organizations will have the option to join the SRSI as strategic members or as associate members. Strategic members will be allocated a seat on the board of the SRSI while associate membership will be required in order to participate in SRSI-hosted projects. SRSI will leverage these membership fees from the supporting organizations to support the transitional activities and operation of the SRSI for an initial period, and will use future membership fees to partially sustain the SRSI in the long term.

---

[3] The actual name used for the community program(s) would be decided by the ECP-ST community

## 4.2 Sustainability Funding: SRSI Foundation

DOE will also provide an initial grant of sustainability funding to the SRSI Foundation, and then subsequent grants based on the level of support it wishes to provide to the ECP-ST product portfolio. The SRSI Foundation will establish a fund for the initial DOE grant and convene an evaluation committee drawn from representatives of the ECP-ST community program(s). The evaluation committee will be the decision making body that determines: a) how funds will be allocated to projects across the community program(s); b) the level of funding required to sustain the existing projects deemed as essential to the community; and c) evaluate the sustainability scorecards for each of the participating projects. Funds will be distributed to the contributing organizations and individuals on this basis.

In the SRSI Model, sustainability scorecards are used to inform funding decisions. Funding is primarily provided to contributors to individual projects in order for them to undertake technical and non-technical work that will improve their project scorecards. This creates a close connection between the projects, which are incentivised to improve their scorecards and which in turn helps the SRSI Foundation to make better overall funding decisions. The SRSI after this transition is shown in Figure 2.

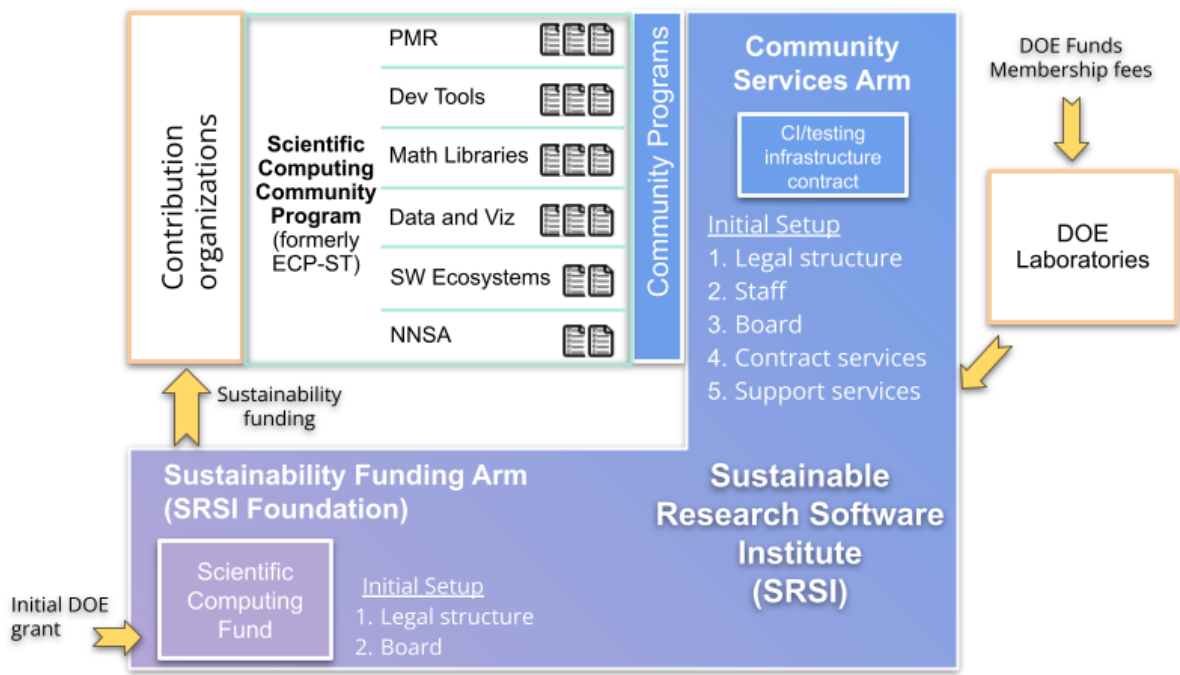

**Figure 2:** *Example SRSI structure that transitions ECP-ST into a "Scientific Computing" community program including an initial contract to provide infrastructure support (likely with Paratools). An initial fund is established in the SRSI Foundation to fund sustainability efforts. DOE provides membership funds to Laboratories which join the SRSI as strategic members.*

## 5. Advantages

In the SRSI Model, the community programs are the locus for in-depth knowledge about their ecosystems. In the case of ECP-ST, detailed knowledge of the project interactions, requirements, dependencies, infrastructure needs, and vendor relationships is an inherent part of the existing leadership and project structure. As the transition process unfolds, SRSI would work closely with the leadership to ensure that this knowledge is preserved during incremental changes to the governance of ST projects.

As the organization grows, participation and collaboration will expand to other agencies and organizations (national and international) sustaining software beyond scientific computing. This vastly broader landscape of communities working on research software will be included in the organization to contribute to creating a more sustainable ecosystem. Additionally, the foundation model potentially simplifies interactions with stakeholder organizations and inter-agency cooperations through the ability to consider novel business arrangements such as the development of a Government Working Capital Fund. Therefore, while embracing the ECP-ST projects during the initial transition phase, our model has the flexibility and inherent ability to scale much more broadly. We would extend an invitation to projects to join as community programs, and any existing ECP-ST projects that better align with these communities could be transitioned.

## 6. Conclusion

This paper describes a transitional model that will lead to the creation of an organization dedicated to the sustainability of research software, while also providing a transition plan for the ECP-ST focus area at the completion of technical activities for ECP. We believe that this will provide the best option for ensuring the continuity of critical exascale software while at the same time establishing a long term plan for sustaining the broader research software ecosystem.

## Acknowledgement

This manuscript has been authored by UT-Battelle, LLC under Contract No. DEAC05-00OR22725 with the U.S. Department of Energy. The United States Government retains and the publisher, by accepting the article for publication, acknowledges that the United States Government retains a nonexclusive, paid-up, irrevocable, world-wide license to publish or reproduce the published form of this manuscript, or allow others to do so, for United States Government purposes. The Department of Energy will provide public access to these results of federally sponsored research in accordance with the DOE Public Access Plan (http://energy.gov/downloads/ doe-public-access-plan).